\documentclass[onecolumn,apjl]{emulateapj}

\def\mbf#1{\mbox{\boldmath ${#1}$}}
\def\Alfven{Alfv\'{e}n~}
\def\Alfvenic{Alfv\'{e}nic~}

\shorttitle{Solar Wind Speeds}
\shortauthors{T. K. Suzuki}

\begin{document}

\title{Forecasting Solar Wind Speeds}

\author{Takeru K. Suzuki$^{1}$}

\email{stakeru@scphys.kyoto-u.ac.jp}
\altaffiltext{1}{Department of Physics, Kyoto University,
Kitashirakawa-Oiwake-cho, Sakyo-ku, Kyoto, Japan 606-8502; 
JSPS Research Fellow}

\begin{abstract}
By explicitly taking into account effects of \Alfven waves, 
I derive from a simple energetics argument a fundamental relation which 
predicts solar wind (SW) speeds in the 
vicinity of the earth from physical properties on the sun. 
Kojima et al. recently found from their observations 
that a ratio of surface magnetic field strength to an expansion factor of 
open magnetic flux tubes is a good indicator of the SW speed. 
I show by using the derived relation that this nice correlation is an 
evidence of the \Alfven wave which accelerates SW in expanding flux tubes. 
The observations further require that fluctuation amplitudes of magnetic 
field lines at the surface should be almost universal in different coronal 
holes, which needs to be tested by future observations.  
\end{abstract}

\section{Introduction}
Speeds of the solar wind (SW) in the vicinity of the earth 
vary from $\sim$300 to $\sim$800km/s \citep{phi95}. 
The SW speed is one of the important parameters 
to predict geomagnetic storms  
triggered by interactions between the SW plasma and the earth 
magnetosphere (e.g. Wu \& Lepping 2002).  
If we can tell SW conditions near the earth 
from observed properties on the sun, we can forecast 
geomagnetic conditions beforehand since it takes a few days until the SW 
reaches us after emanating from the sun. 

Thus, various attempts have been carried out to derive simple 
relations which connect physical properties on the sun and SW speeds 
near the earth. 
Wang \& Sheeley (1990; 1991, hereafter WS90 and WS91) showed that SW speeds 
are anti-correlated with 
expansion factors of magnetic flux tubes from their long-term observations 
as well as by a simple theoretical model, and this relation is widely used 
to predict SW speeds (e.g. Arge \& Pizzo 2000).  
Fisk, Schwadron, \& Zurbuchen (1999; hereafter FSZ) claimed that the SW speed 
should have a positive dependence on the magnetic field strength on the sun. 
\citet{sm03} puts forward a SW scaling which explains the observed 
anti-correlation between the SW speed and freezing-in temperatures, 
reflecting the coronal temperature, of ions \citep{gei95}. 
\citet{ml05} further introduced a correlation between a scale length in the 
chromosphere and the SW speed.      

Turning to the acceleration mechanism of the SW, it is generally 
believed that the \Alfven wave is a promising candidate which dominantly 
works both in heating and accelerating the SW plasma (Belcher 1971; 
Ofman 2004, Cranmer 2005; Suzuki \& Inutsuka 2005; 2006; hereafter, SI05 and 
SI06).
However, there is no fundamental relation derived so far, which is directly 
linked with the \Alfven wave. 
The aim of the present paper is to derive a simple formula which connects 
the SW speed and the solar surface conditions through \Alfven waves 
by referring to results of recent numerical simulations (SI05, SI06). 

\begin{figure}[b]
\figurenum{1} 
\epsscale{0.8}
\plotone{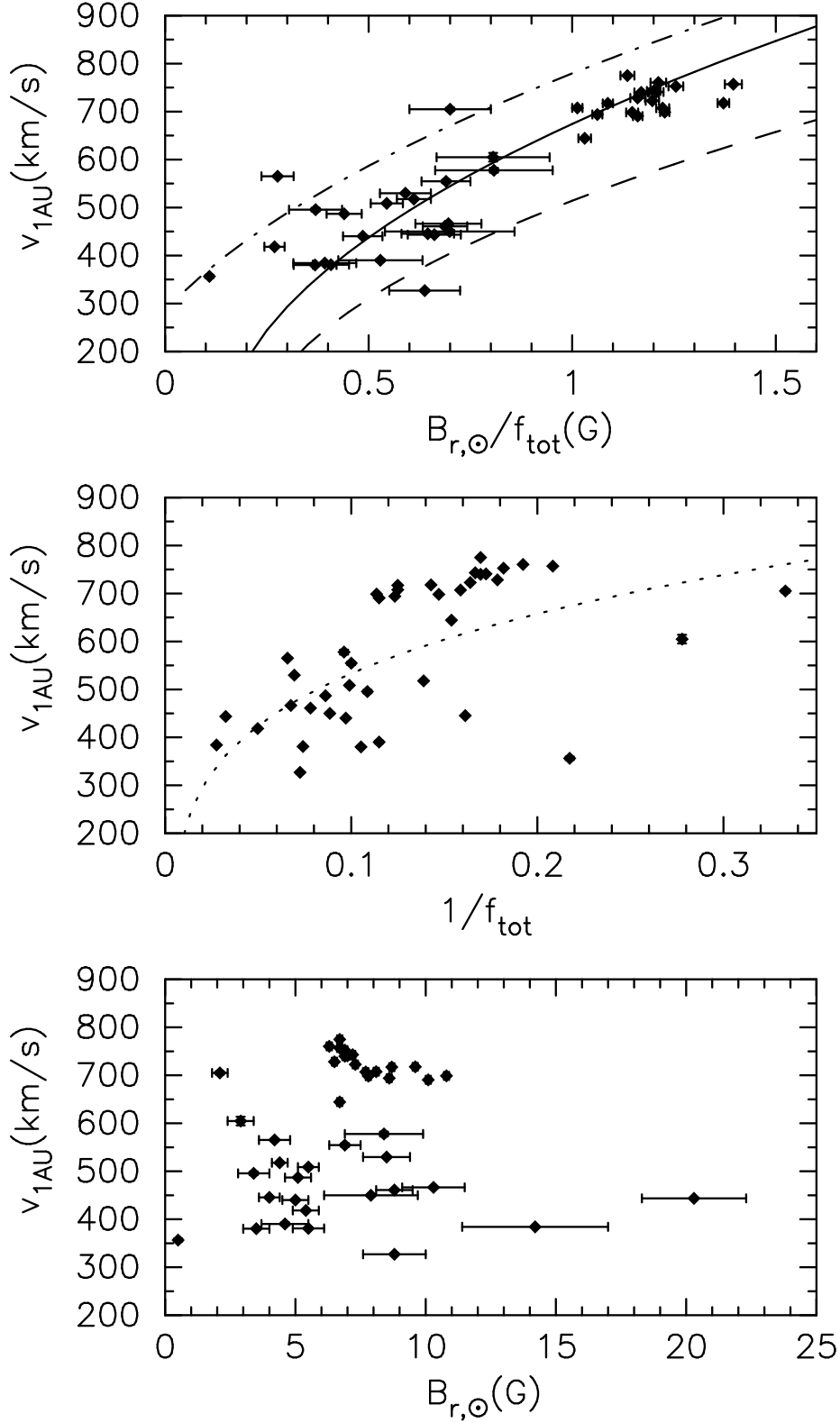} 
\caption{Relations between SW speeds at $r\simeq 1$AU, $v_{\rm 1AU}$, and 
properties of magnetic flux tubes. 
Observed data are from Kojima et al. (2005).  
Coronal magnetic fields are extrapolated from $B_{r,\odot}$ by 
the potential field-source surface method \citep{hk99}. 
$f_{\rm tot}$ is derived from comparison between the areas of open coronal 
holes at the photosphere and at the source surface ($r=2.5R_{\odot}$).  
$v_{\rm 1AU}$ is obtained by interplanetary scintillation 
measurements.  
$v_{\rm 1AU}$, $B_{r,{\odot}}$, and $f_{\rm tot}$ are
averaged over the area of each coronal hole and the data points 
correspond to individual coronal holes.  
(Top) : $v_{\rm 1AU}$ on $B_{r,\odot}/f_{\rm max}$. Lines are 
theoretical prediction from Equation (\ref{eq:swv}). Solid line indicates  
the fiducial case ($\langle \delta B_{\perp} \delta v_{\perp} \rangle 
=8.3 \times 10^5$G cm s$^{-1}$ and $T_{\rm C}=10^6$K). Dot-dashed line 
adopt higher coronal temperature ($T_{\rm C}=1.5\times 10^6$K) with the 
fiducial  $\langle \delta B_{\perp} \delta v_{\perp} \rangle$. Dashed line 
adopt smaller $\langle \delta B_{\perp} \delta v_{\perp} \rangle$
($=5.3 \times 10^5$G cm s$^{-1}$) with the fiducial temperature. 
(Middle) : The same data are plotted in $1/f_{\rm tot} - v_{\rm 1AU}$ plane. 
Dotted line is the result of Equation (\ref{eq:swv}) adopting the similar 
conditions to those considered in WS91 (see text). 
(Bottom) : The same data are plotted in $B_{r,\odot} - v_{\rm 1AU}$ plane. 
}
\label{fig:taudp}
\end{figure}

From an observational viewpoint Kojima et al.(2005) have extensively 
surveyed relations between 
SW speeds around one astronomical unit (1AU), $v_{\rm 1AU}$, and properties 
of magnetic flux tubes, radial magnetic field strength 
at the photosphere,  $B_{r,\odot}$, and a (total) super-radial expansion 
factor of the tube, $f_{\rm tot}$ during a solar minimum phase, 1995-1996 
(Figure 1), where the values are averaged 
over each open coronal hole and the potential field-source surface 
method (e.g. Sakurai 1982) is used to derive $f_{\rm tot}$.  
They claimed that a ratio, $B_{r,\odot}/f_{\rm tot}$, is the best 
indicator of $v_{\rm 1AU}$ (top panel) (see also Suess et al.1984), 
whereas they also found a moderate correlation of 
$v_{\rm 1AU}-1/f_{\rm tot}$ (middle panel) (WS90) and a weak correlation 
of $v_{\rm 1AU}-B_{r,\odot}$ (bottom panel) (FSZ).
Note that, only within the framework of the potential field-source surface 
method, $B_{r,\odot}/f_{\rm tot}$ is equivalent with magnetic field 
strength at the source surface (assumed at 2.5$R_{\odot}$; $R_{\odot}$ is 
the solar radius), the outside of which field lines are 
assumed be radially oriented \citep{hk99}, while they use the ratio of 
$B_{r,\odot}$ and $f_{\rm tot}$ as it stands because it is more 
physically motivated (see \S 3). $f_{\rm tot}$ used in this letter is defined 
as the total expansion factor from the solar surface to 1AU.   

The obtained nice correlation of $v_{\rm 1AU}-B_{r,\odot}/f_{\rm tot}$ seems 
quite reasonable; the positive correlation on $B_{r,\odot}$ appears natural 
since $B_{r,\odot}$ controls strength of Poynting energy which is injected 
from the surface and finally accelerates the SW (FSZ); the negative 
dependence on $f_{\rm tot}$ seems reasonable as well because $f_{\rm tot}$
determines adiabatic loss of the SW in the flux tubes (WS91) 
(Figure 2).   
One may further speculate that the $v_{\rm 1AU}-B_{r,\odot}/f_{\rm tot}$ 
relation reflects \Alfven waves, a type of Poynting flux, 
in the diverging flux tubes.
Here I develop this consideration to give a quantitate interpretation 
of the relation.

\begin{figure}[b]
\figurenum{2} 
\epsscale{0.6}
\plotone{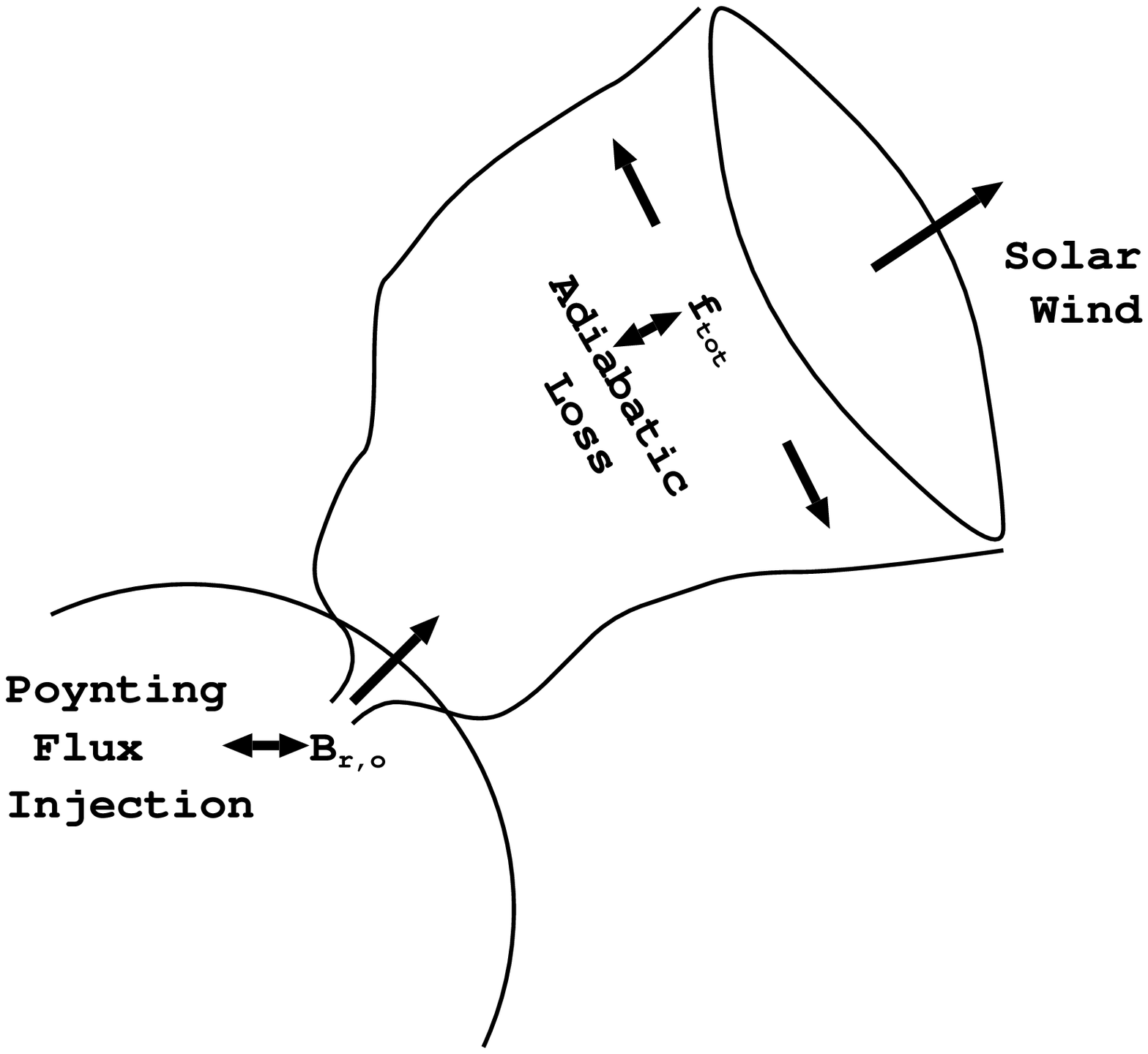} 
\caption{Schematic picture of SW in a magnetic flux tube 
which is super-radially open. 
$B_{r,\odot}$ is proportional to 
Poynting flux input from the surface. $f_{\rm tot}$ determines 
adiabatic loss in the flux tube. Therefore, the kinetic energy of the SW 
in the outer region is inferred to have positive dependence on 
$B_{r,\odot}$ and negative dependence on $f_{\rm tot}$.}
\end{figure} 

\newpage
\section{Formulation}
I derive a simple relation which determines the SW speed near the earth from 
conditions on the solar surface based on a basic energy conservation relation. 
I consider an open magnetic flux tube measured by heliocentric distance, 
$r$, which is anchored at the solar surface, $r=R_\odot$. 
Under the steady-state condition, the energy equation becomes
\begin{equation}
\label{eq:engeq1}
\mbf{\nabla \cdot}\left[\rho \mbf{v} (\frac{v^2}{2} + \frac{\gamma}
{\gamma -1}RT - \frac{GM_{\odot}}{r}) - \frac{1}{4\pi}\mbf{(v\times
B)\times B} +\mbf{F_{\rm c}}\right] + q_{\rm R} = 0 ,
\end{equation}
where $\rho$, $\mbf{v}$, $T$, $\mbf{B}$
$\mbf{F_c}$ and $q_{\rm R}$ are 
density, velocity, temperature, magnetic field strength, 
conductive flux and radiative cooling, respectively. 
$R$, $\gamma$, $G$ and $M_{\odot}$ are respectively
gas constant, a ratio of specific heat, the gravitational constant and 
the solar mass. 
The term involving $\mbf{B}$ denotes Poynting flux under the ideal 
magnetohydrodynamical approximation. 
The cross section of the tube is assumed to expand in proportion to 
$r^2 f(r)$, where $f(r)$ is a super-radial expansion function ($f({R_\odot})
=1$ and $f(r_{1{\rm AU}})=f_{\rm tot}$)\citep{kh76}. 
Note that divergence of an arbitrary vector, 
$\mbf{A}$, becomes $\mbf{\nabla \cdot A} = \frac{1}{f r^2}\frac{d}{dr}
(f r^2 A_r)$. 

The Poynting flux term can be divided into two parts,  
$-\frac{1}{4\pi} (\mbf{v \times B}) 
\times \mbf{B} = \frac{1}{4\pi} (- B_r \delta B_{\perp} \delta v_{\perp} 
+ v_r \delta B_{\perp}^2)$, 
where subscript $r$ denotes the component along the flux tube and $\perp$ 
indicates the tangential components; $\delta B_{\perp}$ and $\delta v_{\perp}$ 
are amplitudes of transverse fluctuations of magnetic field and velocity. 
The first term indicates shear of magnetic field, corresponding to \Alfven 
waves, and the second term denotes advection of magnetic energy. 

Following FSZ, I consider the energy conservation in the flux tube between 
at the solar surface ($r=R_\odot$) and at 1AU ($r=215R_\odot$).  
At the surface, besides the gravitational potential, dynamical and magnetical 
energy associated with the surface convection gives an important contribution. 
Here I rearrange these terms concerning the inputs of energy by the convection 
into two parts, 
$ - \frac{1}{4\pi}[\mbf{(v\times B)\times B}]_r + \rho v(\frac{v^2}{2} + 
\frac{\gamma}{\gamma-1}RT)= -\frac{1}{4\pi} B_r \delta B_{\perp} 
\delta v_{\perp} 
+F_{\rm H}$, namely incompressive part (\Alfven wave), $-\frac{1}{4\pi} 
B_r \delta B_{\perp} \delta v_{\perp} $, and compressive part, 
$F_{\rm H}$. Note that the magnetic energy term ($v_r\delta B_{\perp}^2/4\pi$) 
is included in $F_{\rm H}$. At 1AU 
the dominant term is the kinetic energy of SW (FSZ).  
Then, the energy conservation in the flux 
tube gives    
\begin{equation}
\label{eq:engeq2}
\left[\rho v r^2 f_{\rm tot} \frac{v^2}{2}\right]_{r={\rm 1AU}} = \left[r^2 
\left(-\frac{B_r \langle 
\delta B_{\perp} \delta v_{\perp}\rangle}{4\pi} +F_{\rm H} - 
\rho v \frac{G M_{\odot}}{r}\right)\right]_{r= R_{\odot}} 
- \int_{R_{\odot}}^{\rm 1AU} dr r^2 
f q_{\rm R} , 
\end{equation}
where $\langle \cdots \rangle$ denotes time-average. Thermal conduction 
does not appear explicitly because it only works in redistribution of 
the temperature structure between the two locations. 

The reason of the decomposition of the energy injection terms into the two 
parts is that their dissipation characters are different. 
Generally, the dissipation of the \Alfven wave is slow since it is hardly  
steepen to shocks. Therefore, \Alfven waves propagate a long distance to 
contribute to the heating and acceleration of the SW around $\sim$ a few to 
$\sim 10R_{\odot}$ (SI06).  
On the other hand, compressive waves and turbulences denoted by 
$F_{\rm H}$ are more dissipative so that they only contribute to the 
heating in the chromosphere \citep{cs92} and the low corona \citep{suz02}.
Most of the energy of $F_{\rm H}$ which dissipates in the corona 
is lost by downward thermal conduction toward the chromosphere, which 
finally radiates away in the transition region and 
the upper chromosphere \citep{hm82a}. The rest ($\sim 10\%$) of the energy is 
transferred to enthalpy flux \citep{wtb88} to keep the hot corona with 
$T\gtrsim 10^6$K.    
The radiative cooling (last term in Equation \ref{eq:engeq2}) is also only 
efficient in the low corona or below where the density is sufficiently high. 
Therefore, subtraction of the radiative loss from $F_{\rm H}$  
gives `effective' coronal temperature, $T_{\rm C}$, namely $F_{\rm H} - 
\frac{1}{R_{\odot}^2}\int_{R_{\odot}}^{1{\rm AU}} 
dr r^2 f q_{\rm R} \approx \rho v \frac{\gamma}{\gamma-1}R T_{\rm C}$. 

Finally, we have a conservation equation : 
\begin{equation}
\left\{\rho v r^2 f_{\rm tot} \frac{v^2}{2}\right\}_{r={\rm 1AU}} = 
\left\{r^2 \left[-\frac{B_r \langle \delta B_{\perp} 
\delta v_{\perp}\rangle}{4\pi} + \rho v \left(\frac{\gamma}{\gamma-1}R 
T_{\rm C} - \frac{G M_{\odot}}{r}\right)\right]\right\}_{r=R_{\odot}}.  
\label{eq:cmp}
\end{equation}
The second term ($\rho v r^2 \frac{\gamma}{\gamma-1}RT_{\rm C}$) 
on the right hand side is evaluated at 
the base of the corona in the strict sense since it implicitly considers  
the energy balance at the transition region (see \S 3). 
However, I use the location, 
$r=R_{\odot}$, because the distance between the photosphere and the corona 
is much smaller than $R_\odot$. 
A conceptional novelty of the present formulation is that I explicitly 
include the \Alfven wave term which was neglected (FSZ) or parameterized 
in more phenomenological ways (WS91; Schwadron \& McComas 2003) 
in previous works. 
Note that $-\delta B_{\perp} \delta v_{\perp}$($>0$ for outgoing \Alfven 
waves) is a conserved quantity of 
\Alfven waves which propagate in static media if they do not dissipate\footnote
{This is derived from conservation of the wave energy flux, 
$
0 = \frac{1}{f r^2}\frac{d}{dr}(f r^2 B_r \delta B_{\perp} \delta v_{\perp})
= B_r \frac{d}{dr}(\delta B_{\perp} \delta v_{\perp}), 
$
where we have used conservation of magnetic field, $B f r^2 = $const. 
Incidently, in moving media wave action should be used as a 
conserved quantity instead of energy flux \citep{jaq77}.}. 
Therefore, $- \langle \delta B_{\perp} \delta v_{\perp}\rangle$ is 
a measure of dissipation and reflection of \Alfven waves in the chromosphere 
and the low corona where the gas is almost static, and the results of 
numerical simulations (SI05; SI06) can be used for the \Alfven wave term in 
a straightforward manner. 
The physical meaning of Equation (\ref{eq:cmp}) is clear; the kinetic energy 
of the SW at 1AU is determined by positive contributions from the input  
\Alfven wave energy (first term) and the thermal pressure 
of the corona (second term) and a negative contribution due to the 
gravitational potential well (third term).

Rearranging Equation (\ref{eq:cmp}) by using the mass conservation relation, 
$[\rho v r^2 f_{\rm tot}]_{r={\rm 1AU}} = [\rho v r^2]_{r=R_\odot}$,  
we can derive a more direct form  
which predicts SW speeds in the vicinity of the earth from the physical 
conditions on the solar surface : 
\begin{eqnarray}
\label{eq:swv}
& &v_{\rm 1AU} = \sqrt{2\times \left(-
\frac{R_{\odot}^2}{4\pi(\rho v r^2)_{\rm 1AU}}\frac{B_{r,{\odot}}}{f_{\rm tot}}
\langle \delta B_{\perp} \delta v_{\perp}\rangle_{\odot} 
+ \frac{\gamma}{\gamma-1}R T_{\rm C} - \frac{G M_{\odot}}{R_{\odot}}
\right)} \\
&=& 
300({\rm km/s})\sqrt{5.9\left(\frac{-\langle \delta B_{\perp}\delta 
v_{\perp}\rangle_\odot}{8.3\times 10^5({\rm cm\;s^{-1}G})}\right)
\left(\frac{B_{r,{\odot}}{\rm (G)}}
{f_{\rm tot}}\right) +3.4 \left(\frac{\gamma}{1.1}\right)\left(
\frac{0.1}{\gamma-1}\right)\left(\frac{T_{\rm C}}{10^6{\rm (K)}}
\right) - 4.2} \nonumber,  
\end{eqnarray}
where I use the observed mass flux at 1AU, $(\rho v)_{\rm 1AU}=5.4\times 
10^{-16}$(g cm$^{-2}$), which 
is almost constant even in SWs with different speeds (e.g. Aschwanden, Poland, 
\& Rabin 2001). 
The velocity amplitude at the surface can be estimated from observed  
granulation motions at the photosphere as $\delta v_{\perp} \simeq 1$km/s 
\citep{hgr78}. 
The magnetic amplitude is derived from $\delta v_{\perp}$ and the 
photospheric density $\rho\simeq 10^{-7}$g cm$^{-3}$ as 
$\delta B_{\perp} = -\delta v_{\perp}\sqrt{4\pi \rho} \simeq -110$G. 
Then, $-\langle \delta B_{\perp} \delta v_{\perp}\rangle_\odot 
\simeq 5.5 \times 10^6$(G cm s$^{-1}$), where a factor of 1/2 is included 
due to the time-average. 
This value is for the case in which all the \Alfven waves from the 
surface propagate into the SW region.  
In the real situation, they suffer reflection in the chromosphere. 
SI05 shows only $\simeq 15$\% of the initial energy propagates outwardly 
to contributes to the heating of the coronal and SW plasma. 
Therefore, we adopt $-\langle \delta B_{\perp} \delta v_{\perp}\rangle_\odot 
\simeq 8.3 \times 10^5$(G cm s$^{-1}$) as a fiducial value. 
As for the thermal pressure, I assume $T_{\rm C}=10^6$K as a typical 
coronal temperature. $\gamma$ should be larger than the adiabatic value 
($\gamma=5/3$) because of the thermal conduction \citep{sue77}; 
I consider $\gamma=1.1$ in this paper. 

\section{Results and Discussions}
I plot relations derived from Equation (\ref{eq:swv}) in the top panel of 
Figure 1. The fiducial case with $\langle \delta B_{\perp} 
\delta v_{\perp}\rangle=8.3\times 10^5$(G cm s$^{-1}$) and $T_{\rm C}=10^6$(K) 
(solid line) explains the observed trend quite well. 
The $v_{\rm 1AU} - B_{r,{\odot}}/f_{\rm tot}$ relation reflects  
the \Alfven waves which accelerate the SW in expanding magnetic 
flux tubes. $B_{r,{\odot}}$ determines energy flux 
of the \Alfven waves ($\propto [B_r\delta B_{\perp} 
\delta v_{\perp}]_{\odot}$). 
$f_{\rm tot}$ controls 'dilution' of the energy flux; in a 
flow tube with larger $f_{\rm tot}$ more energy is used to expand the tube 
rather than transfered to the kinetic energy of SW.  
Thus, the positive dependence on $B_{r,\odot}$ and the negative dependence 
on $f_{\rm tot}$ are naturally derived. 
The result does not depend on different dissipation processes of \Alfven 
waves because I only consider the SW speed at the sufficiently distant 
location ($r=1$AU) where the wave energy are mostly dissipated.  

The top panel of Figure 1 also exhibits that  
$B_{r,\odot}/f_{\rm tot}$ is the most important parameter in determining 
the SW speed and that other physical conditions on the solar surface should 
be similar. 
Almost all the data are between dot-dashed and dashed lines which are 
the results of cases adopting slightly larger $T_{\rm C}$($= 1.5\times 
10^6$K) and smaller $\langle \delta B_{\perp} \delta v_{\perp}\rangle$
($=5.3\times 10^5$G cm s$^{-1}$) than the fiducial case.  
Particularly, the difference of the wave amplitudes ($\delta v_{\perp}
\propto \sqrt{\langle \delta B_{\perp} \delta v_{\perp}\rangle}$) 
between the solid and dashed lines are only 20\%.         
This indicates that the amplitudes of \Alfven 
waves at the surface should be very similar in different coronal holes.  
The observed data are from not only polar coronal holes but mid-latitude and 
equatorial coronal holes, some of which are located near active regions. 
Therefore, one can infer that the amplitudes could vary a lot 
since the circumstances are quite different.  However, the observation 
seems to favor the constancy of the \Alfvenic fluctuations at the 
footpoints. 
Although it is very difficult to observe \Alfvenic motions 
of field lines on the solar surface at present \citep{ul96}, 
this can be observationally studied in the very near future by 
Solar-B satellite which can stably observe fine-scale motions of surface 
magnetic fields. 

Let me compare the present analysis with a model calculation for the 
$v_{\rm 1AU}-1/f_{\rm tot}$ correlation by WS91 (see also \citet{w93}). 
Although it is not simple to compare both since the assumptions adopted 
in WS91 are different from mine (for example, WS91 fixed the coronal base 
density, while I adopt the constant mass flux at 1AU),  
I can derive a relation of $v_{\rm 1AU}-1/f_{\rm tot}$ from 
Equation (\ref{eq:swv}) by using similar constraints to those considered 
in WS91.   They adopted 
a constant field strength at 1AU, $B_{r,{\rm 1AU}}=3\times 10^{-5}$G, and 
assumed a constant energy flux ($= 1.5\times 10^5$erg cm$^{-2}$s$^{-1}$) 
at the coronal base. 
In the middle panel of Figure 1, I present the result with fixed 
$B_{r,\odot}/f_{\rm tot}=1.4$, corresponding to 
$B_{r,\odot}=3\times 10^{-5}$G, and energy flux of \Alfven waves, 
$B_{r,\odot}\langle \delta B_{\perp} \delta v_{\perp}\rangle/4\pi = 
4\times 10^5$erg cm$^{-2}$s$^{-1}$ ({\it i.e.} $\langle \delta B_{\perp} 
\delta v_{\perp}\rangle \propto f_{\rm tot}^{-1}$) in Equation (\ref{eq:swv})
by the dotted line. Note that I need the larger energy flux because 
the inner boundary is not the coronal base but the photosphere.
The figure shows that the dotted line follows the average trend of the data 
and the result of WS91 is reasonable.  
However, I would like to address that 
some data are located away from the main $v_{\rm 1AU}-1/f_{\rm tot}$ trend 
and they can be explained in a unified manner by taking into account 
$B_{r,\odot}$. 

The relation of Equation (\ref{eq:swv}) seemingly contradicts to the reported 
anti-correlation of the coronal temperature and the SW speed \citep{gei95}. 
This is because my treatment of 
the thermal processes near the surface is too much simplified; the complicated 
energy balance from the chromosphere to the low corona is represented only 
by the `effective' enthalpy, 
$\frac{\gamma}{\gamma-1}R T_{\rm C}$. The formulation for the 
temperature-velocity relation by \citet{sm03} 
is complementary to the present formulation.  In \citet{sm03} the 
detailed energy balance at the transition region is taken into account, 
while they assumed a constant input of the \Alfven wave energy flux which I 
investigate more in detail.    

In this letter, in order to focus on the SW speed, I simply apply the 
observed (almost) constant mass flux at 1AU when deriving the relation 
for the SW speed. 
For self-consistent treatments, however, it is important to study how to 
determine the mass flux of the SW not only by numerical simulations 
(e.g. SI06) but by simple models.

I think that Equation (\ref{eq:swv}) is applicable to 
SWs during both sunspot minimum and maximum phases because it is derived 
based only on the simple energetics.  
At present, however, the observed data (Kojima et al.2005) which I use for the 
comparison are only during the sunspot minimum phase (1995-1996).   
In order to study the generality of the derived relation, comparisons with 
SW data during different phases (Fujiki et al.2006) are important.   
One should be careful that $f_{\rm tot}$ which should be used for the 
prediction is the actual super-radial expansion 
factor from $R_{\odot}$ to 1AU, while in most cases, including Kojima et al.
(2005),  $f_{\rm tot}$ is observationally estimated from the 
comparison between at $R_{\odot}$ and at the source surface ($2.5R_{\odot}$) 
based on the potential field-source surface method. 
Errors due to this method could be non-negligible if the potential 
approximation becomes worse and/or if one separately treats flux tubes 
in a coronal hole (WS90). (In this sense, Kojima et al.2005 as well as I use 
$B_{r,\odot}/f_{\rm tot}$ instead of field strength at the source surface.) 
Thus, the precise determination of coronal magnetic fields (e.g. 
Linker et al.1999) is important for reliable forecasts of SW 
speeds from the relation of Equation (\ref{eq:swv}).

The author thank the solar wind group in Nagoya-STEL 
(Kojima, M., Tokumaru, M., Fujiki, K.) for fruitful discussions and 
providing observational data. The author also thank the referee for 
constructive comments.    
This work is supported by the JSPS Research Fellowship for Young
Scientists, grant 4607.

\end{document}